
\magnification=\magstep1
\centerline {\bf REPRESENTATIONS OF THE  $U_q(u_{4,1})$ AND A $q$-POLYNOMIAL}
\centerline {\bf THAT DETERMINES BARYON MASS SUM RULES}
\vskip 30 pt
\centerline {\bf A.M.Gavrilik,\ \ I.I.Kachurik,\ \ A.V.Tertychnyj}
\vskip 20 pt
\centerline {Institute for Theoretical Physics of}
\centerline {National Academy of Sciences of Ukraine}
\centerline {252143  Kiev-143, Ukraine}
\vskip 40 pt
\noindent  {\bf Abstract.}  With quantum groups $U_q(su_n)$ taken as
classifying symmetries for hadrons of $n$ flavors, we calculate
within irreducible representation $D^+_{12}(p-1,p-3,p-4;p,p-2)$
\ ($p \in {\bf Z}$) of 'dynamical' quantum group $U_q(u_{4,1})$
the masses of baryons ${1\over 2}^+$ that belong to ${\it 20}$-plet
of $U_q(su_4)$. The obtained $q$-analog of mass relation (MR) for
$U_q(su_3)$-octet contains unexpected mass-dependent term multiplied
by the factor ${A_q\over B_q}$ where $A_q,\  B_q$ are certain polynomials
(resp. of 7-th and 6-th order) in the variable $q+q^{-1}\equiv [2]_q$. Both
values $q=1$ and $q=e^{i\pi \over 6}$ turn the polynomial $A_q$ into zero.
But, while $q=1$ results in well-known Gell-Mann--Okubo (GMO) baryon MR,
the second root of $A_q$ reduces the $q$-MR to some novel mass sum rule
which has irrational coefficients and which holds, for empirical masses,
even with better accuracy than GMO mass sum rule.
\vskip 50 pt
\centerline {\bf 1. INTRODUCTION}
\bigskip
 Applications of quantum algebras ($su_q(2)$ first of all) to
phenomenological description of rotational spectra of deformed heavy nuclei
and diatomic molecules, have appeared a couple of years ago and seem to be
encouraging [1-3] (concerning physical applications of quantum
groups/algebras in a wider context see [4] and references therein).

Recently, the use of higher rank quantum algebras $su_q(n)$
in order to replace conventional algebras $su(n)$ of unitary groups
and their irreducible representations (irreps) in describing global
symmetries of hadrons (vector mesons) of $n$ flavours
has been proposed [5]. With the help of the corresponding
algebras $u_q(n+1)$ of 'dynamical' symmetry, one can realize
necessary breaking of flavor symmetries up to exact (for strong
interactions alone) isospin symmetry $su_q(2)_I$ and obtain some
$q$-analogs of mass relations (MR's). It was demonstrated that at
every fixed $n$,\ $n=3,...,6$,\ all the $q$-dependence in vector
meson masses and in coefficients of their $q$-analog MR's
appears, modulo possible additional $q$-deuce $[2]_q$,
\ only through the ratio ${[n]_q\over [n\!-\!1]_q}$.
\ By means of the relation
$${[n]_q\over [n-1]_q }=
1+{[n]_q-[n\!-\!1]_q\over [n-1]_q}=1+{\Delta_{\{(2n-1)_1\}}\over
1+ \sum_{r=2}^{n-1} \Delta_{\{(2r-1)_1\}} }, $$
where the Lorant-type polynomials $[n]_q - [n\!-\!1]_q$\ of \ the
$q$ \ coincide (at least formally) with the Alexander polynomials
 $\Delta_{\{(2n-1)_1\}}(q)$\ corresponding to the toroidal knots
$(2n-1)_1$, the $q$-dependences in masses and in $q$-MR's were
shown to be expressible [6] completely in terms of these
knot invariants.

A comparison with empirical data requires appropriate fixation of
deformation parameter, and it appears that to every number of
flavors $n,\ n\ge 3,$ there corresponds a prime root of unity
$q=q(n)=e^{{i\pi}/{(2n-1)}}$. The latter turns into zero
the polynomial $[n]_q-[n-1]_q$ (equivalently, respective Alexander
polynomial of the toroidal $(2n-1)_1$-knot).
In a sense, the polynomial $[n]_q-[n-1]_q$ through
its root $q(n)$ determines the strength of deformation at every
fixed $n$, and due this property may be called a
{\it defining polynomial} for the corresponding vector meson mass
sum rule (MSR).

Utilizing the {\it quantum} algebras instead of
conventional unitary groups of flavor symmetries,
together with 'dynamical' {\it quantum} algebras,
we get as a result that the collection of torus knots
$5_1,\ 7_1,\ 9_1,\ 11_1$ \ is put into
correspondence [6] with vector quarkonia
$s\bar s$, $c\bar c$, $b\bar b$, and $t\bar t$ respectively.
Thus, application of the embedding
$U_q(u_{n})\subset U_q(u_{n+1})$ to vector meson masses
provides an appealing possibility of certain {\it topological
characterization of flavors}, since the number $n$ just corresponds
to  $2n\!-\!1$  overcrossings of 2-strand braids whose closures give
these $(2n-1)$-torus knots. Equivalently, using $(a,b)$-presentation
of these same knots with $a=2n-1,\ b=2$, one is led to the
correspondence: $n \leftarrow\!\rightarrow w\equiv 2n-1$,\ where
$w$ (or $a$) is nothing but the winding number around the body (tube)
of torus (winding number around the hole of torus being equal to 2
for all $n\ge 3$).

Our present goal is to extend the approach of [5,6] to the case
of baryons ${1\over 2}^+$ (including charmed ones) again adopting
$U_q(u_4)$ for the 4-flavor symmetry. However, like in the situation
of baryon MSR's obtained with non-deformed dynamical pseudounitary
$u(4,1)$-symmetry [7,8], it will be more convenient for us now to
exploit representations of 'noncompact' dynamical symmetry,
realized by the quantum algebra $U_q(u_{4,1})$, in order
to effect necessary symmetry breakings.

The paper is organized in the following manner. In Sec.2 and Sec.3,
certain amount of information concerning (universal enveloping)
quantum algebras $U_q(u_n)$,\  $U_q(u_{n,1})$\ as well as their
finite- and infinite-dimensional representations necessary
for the considered application, is presented.

On the base of calculation of baryon masses (Sec.4) within
concrete representation $D^+_{12}(p-1,p-\!3,p-4;p,p-2)$
\ (where $p$ is some fixed integer, precise value of which is
unessential since it will not enter final expressions for masses),
it is demonstrated in Sec.5 that the resulting $q$-analog of
baryon octet MR takes somewhat unusual form since it contains
an additional mass-dependent structure multiplied by the ratio
${A_q\over B_q}$ of certain $q$-polynomials. The ('rigid')
$q$-fixation procedure described in the 6th Section leads to
conclusion that the $q$-analog of baryon octet MR yields
either the usual Gell-Mann--Okubo (GMO) mass sum rule [9]
$$m_N+m_{\Xi}={3\over 2}m_{\Lambda}+{1\over 2}m_{\Sigma } \eqno (1)
$$
or a very successful novel MSR with irrational coefficients (see
eq.(20) below) if one fixes the deformation parameter as  $q=1$
\ or  $q=e^{i\pi \over 6}$\ respectively. These values are nothing
but two distinct roots of one and the same {\it defining polynomial}
$A_q$\ appearing in the $q$-analog. With the two alternative choices
of deformation parameter, consequences for masses of charmed
baryons may be obtained and compared to each other.

Another $q$-analog of octet MR (with different defining
$q$-polynomial $\tilde A_q$) is obtained in Sec.7 by
calculations within the other specific representation
$\tilde D^+_{32}(p-1,p-3,p-4;p-4,p-2)$ of $U_q(u_{4,1})$.
It can be shown, however, that from the $q$-dependent expressions
for octet baryon masses obtained by means of this representation
$\tilde D^+_{32}(\ldots )$, the same $q$-analog with same
$q$-polynomial $A_q$ mentioned in the previous paragraph (modulo
minimal modification which consists in the replacement
$B_q\rightarrow\tilde B_q$\ with certain $q$-polynomial
$\tilde B_q$) can also be derived. Last Section is devoted to
some concluding remarks.

\bigskip
\centerline {\bf 2. QUANTUM ALGEBRA $U_q(gl_n)$ AND ITS REAL FORMS}
\bigskip
We will use the denotion
$[B]_q \equiv [B]\equiv (q^B - q^{-B})/(q - q^{-1})$
where $B$ is either a number or an operator. The elements
${\bf 1},\  A_{jj+1},$\  $A_{j+1j},$\  $A_{jj},\ j=
1,2,...,n-1,$\  $A_{nn}$ that generate the $q$-deformed (universal
enveloping) algebra $U_q(gl_n)$, satisfy the relations [10]
$$\eqalignno { &[A_{ii}, A_{jj}] = 0,   \ \ \ \ \ \ \ \ \ \ \
[A_{ii}, A_{jj+1}]={\delta }_{ij}A_{ij+1}-{\delta }_{ij+1}A_{ji}, &{}\cr
&[A_{ii}, A_{j+1j}]={\delta }_{ij+1}A_{ij}-{\delta }_{ij}A_{j+1i}, &{}\cr
&[A_{ii+1}, A_{j+1j}]={\delta }_{ij} [A_{ii} - A_{i+1i+1}]_q, &{}\cr
&[A_{ii+1}, A_{jj+1}]=[A_{i+1i}, A_{j+1j}]=0 \quad \quad
\hbox {for}\quad \quad \vert {i-j}\vert \ge 2,   &{(2)} } $$
and the trilinear ($q$-Serre) relations
$$\eqalignno {(&A_{i\mp 1i})^2A_{ii\pm 1} -
[2]_qA_{i\mp 1i}A_{ii\pm 1}A_{i\mp 1i} +
 A_{ii\pm 1}(A_{i\mp 1i})^2 = 0,  &{}\cr
(&A_{ii\pm 1})^2A_{i\mp 1i} - [2]_qA_{ii\pm 1}A_{i\mp 1i}A_{ii\pm 1} +
 A_{i\mp 1i}(A_{ii\pm 1})^2 = 0.                 &{(3)} }
$$
Endowed with comultiplication, counit and antipode (which we do not
reproduce here), the $q$-deformed algebra $U_q(gl_n)$\ becomes a
quantum (Hopf) algebra.

In what follows, we'll need both compact and non-compact real forms
of  $U_q(gl_n)$.
\medskip
The {\it 'compact' quantum algebra} $U_q(u_n)$
is singled out by means of the *-operation
 $$(A_{jj})^* = A_{jj},\quad \quad \quad
(A_{j+1j})^* = A_{jj+1},\quad \quad \quad
                     (A_{jj+1})^* = A_{j+1j}. \eqno (4)$$
\medskip
The {\it 'noncompact' quantum algebra} $U_q(u_{n,1})$ is singled out
from  $U_q(gl_{n+1})$ by introducing another *-operation which
includes relations (4) of 'maximal compact' subalgebra $U_q(u_n)$
and, in addition, the relations
$$ (A_{n+1n})^* = - A_{nn+1},\ \ \ \ (A_{nn+1})^* = - A_{n+1n},
\ \ \ \ (A_{n+1\ n+1})^* = A_{n+1\ n+1}.       \eqno {(5)} $$
\medskip
Finite-dimensional representations of $U_q(u_n)$, similarly to those
of  the non-deformed algebra $u_n$, are given by sets of ordered
integers  ${\bf m}_n$ = $(m_{1n}, m_{2n},..., m_{nn})$ and, since
standard branching rules survive through $q$-deformation, realized
by means of ($q$-analog of) Gel'fand-Tsetlin basis and formulas.
Representation formulas for $A_{ii}$ remain unchanged, and $A_{kk + 1},
A_{k+1k},\ k=1,...,n-1,$ act according to formulas given in [10].
Action formulas for the operators which represent nonsimple-root
elements must be consistent with $q$-Serre relations (3). We use
$A_{ij}$ for $\vert i-j\vert = 2$ in the form (see e.g. [5,6])
$$\eqalignno {A_{kk+2}&=A_{kk+2}(q)\equiv q^{1/2}A_{kk+1}A_{k+1k+2}-
                q^{-1/2}A_{k+1k+2}A_{kk+1},     &{(6a)}\cr
	    A_{k+2,k}&=A_{k+2,k}(q)\equiv q^{1/2}A_{k+1k}A_{k+2,k+1}-
		q^{-1/2}A_{k+2,k+1}A_{k+1k},      &{(6b)} }$$
such that the $q$-Serre relations corresponding to upper signs in (3)
follow from (6a) and the commutation rules (CR's)
$$\eqalignno {q^{1/2}A_{k+1k+2}A_{kk+2}&-q^{-1/2}
A_{kk+2}A_{k+1k+2}=0, &{}\cr
     q^{1/2}A_{kk+2}A_{kk+1}&-q^{-1/2}A_{kk+1}A_{kk+2} = 0,&{(7a)} }$$
whereas those corresponding to lower signs in (3) follow from (6b) and
the CR's
$$\eqalignno {q^{1/2}A_{k+2,k+1}A_{k+2,k}&-q^{-1/2}
A_{k+2,k}A_{k+2,k+1}=0, &{}\cr
     q^{1/2}A_{k+2,k}A_{k+1k}&-q^{-1/2}A_{k+1k}A_{k+2,k} = 0.&{(7b)} }$$
Dual definition $\tilde A_{kk+2}\equiv -A_{kk+2}(q^{-1}),$\ \
 $\tilde A_{k+2,k}\equiv -A_{k+2,k}(q^{-1})$ \ is paired with respective
dual CR's. Operators $A_{ij}$ for other nonsimple-root elements
($\vert i-j\vert > 2$) are treated analogously.
\bigskip
\centerline {\bf 3. ON THE REPRESENTATIONS OF $U_q(u_{n,1})$ }
\bigskip
In this section we consider some details
concerning irreducible representations (irreps) not just for the
particular case of $U_q(u_{4,1})$, but for more general case
of quantum algebras $U_q(u_{n,1})$,\ $2\le n <\infty$\ (see [11, 12]).
First of all, we have to remark that the construction of the
'principal nonunitary series' of representations, analysis of their
(ir)reducibility as well as the classification of irreps and 'unitary'
(i.e. infinitesimally unitary) irreps of $U_q(u_{n,1})$  runs in much
analogous way to that of the non-deformed (that is, $u(n,1)$) case.
Let us refer to [8, 13] and references given therein for the
non-deformed case.

Throughout this section, $q$ is considered to be generic (not equal
to a root of unity). The representations of the algebra
$U_q(u_{n,1})$ are characterized by their signatures $\chi $,
that is, by the sets of $n+1$ numbers: $\chi \equiv
(l_1,l_2,...,l_{n-1};c_1,c_2)$. Here $c_1, c_2$ are complex
numbers such that $c_1+c_2\in {\bf Z}$, and all the
$l_i, \ i=1,...n-1,$\ are integers related with the components
$m_1,m_2,..,m_{n-1}\equiv {\bf m}$\ of the highest weight
${\bf m}$\ of irrep of the subalgebra $U_q(u_{n-1})$, namely,
$l_i=m_i-i-1$. The condition on the components of highest weight
in terms of $l_i$ reads: $l_1>l_2>...>l_{n-1}$. Under restriction
to the 'compact' subalgebra  $U_q(u_n)$, the representation
$T_{\chi}$ decomposes into direct sum of all those irreps
$T_{{\bf l}_n}$ (${\bf l}_n\equiv (l_{1n},l_{2n},...,l_{nn}),$\
$l_{jn}=m_{jn}-j$,\ $j=1,...,n$,\ where $m_{1n},m_{2n},...,m_{nn}$
form the highest weight ${\bf m}_n$\ of irrep of $U_q(u_n)$) for
which the condition
$$l_{1n}>l_1\ge l_{2n}>l_2\ge ...\ge l_{n-1n}>l_{n-1}\ge l_{nn}
\eqno(8)$$
is satisfied. All representations $T_{{\bf l}_n}$ which satisfy eq.(8)
are contained in $T_{\chi}$ with unit multiplicity.

Action of the representation $T_{\chi}$ is defined in the
carrier Hilbert space taken as a direct sum of finite-dimensional
carrier spaces of irreps of $U_q(u_n)$. In the carrier space
of $T_{\chi}$, we choose a canonical orthonormal basis
formed by the union of canonical (Gel'fand -- Tsetlin) bases
of $T_{{\bf l}_n}$, in accordance with the
reduction chain $U_q(u_{n,1})\supset U_q(u_{n})\supset
U_q(u_{n-1})\supset ...\supset U_q(u_{2})$.
An (orthonormalized) basis vector is completely characterized by the
set $\chi , {\bf l}_{n}$, ${\bf l}_{n-1},...,$ ${\bf l}_2, {\bf l}_1$
(here ${\bf l}_k\equiv (l_{1k}, l_{2k},...,l_{kk})$;\
$l_{ik}\equiv m_{ik}-i,\ i=1,
...,k$)  and will be denoted as
$$\vert \chi; {\bf l}_{n}, {\bf l}_{n-1},..., {\bf l}_1
\rangle .                                     \eqno (9)
$$
When restricted to subalgebra $U_q(u_{n})$, representation
operators act according to formulas of ref. [10]. Operators
$T_{\chi }(A_{nn+1})$ and $T_{\chi }(A_{n+1n})$ that
represent 'noncompact' generators of $U_q(u_{n,1})$ act
according to formulas
$$\eqalignno  { &T_{\chi }(A_{nn+1})
\vert {\bf {\chi}}; {\bf l}_{n}, {\bf l}_{n-1},..., {\bf l}_1\rangle
=       &{}\cr
&{\sum}_{r=1}^{n}\Bigg\vert { [c_1-l_{rn}][l_{rn}-c_2]
\Pi_{j=1}^{n-1}[l_{jn-1}-l_{rn}-1][l_{rn}-l_j]
\over \Pi_{s=1,s\ne r}^n [l_{rn}-l_{sn}+1]
[l_{rn}-l_{sn}]  }\Bigg\vert^{1/2}
\vert {\bf {\chi}} ,{\bf l}_n^{+r},
{\bf l}_{n-1},...,{\bf l}_1\rangle \quad  &{(10)}  }
$$
\noindent {and}
$$\eqalignno  { &T_{\chi }(A_{n+1n})
\vert {\bf\chi}; {\bf l}_{n}, {\bf l}_{n-1},..., {\bf l}_1\rangle
=       &{}\cr
&\sum_{r=1}^{n}\!\Bigg\vert { [c_1-l_{rn}+1][l_{rn}-c_2-1]
\Pi_{j=1}^{n-1}[l_{jn-1}-l_{rn}][l_{rn}-l_j-1]
\over \Pi_{s=1,s\ne r}^n [l_{rn}-l_{sn}]
[l_{rn}-l_{sn}-1]  }\Bigg\vert^{1/2}\!
\vert {\bf\chi} ,{\bf l}_n^{-r},
{\bf l}_{n-1},...,{\bf l}_1\rangle \quad  &{(11)}  }
$$
where ${\bf l}_n^{\pm r}$ means that the component $l_{rn}$
in ${\bf l}_n$ is to be replaced respectively by $l_{rn}\pm 1$.

To have representation formulas for other 'noncompact' operators
$T_{\chi }(A_{jn+1})$ and $T_{\chi }(A_{n+1j})$,\  $1\le j\le n-1$,
one has to utilize relations analogous to eqs. (6).

By means of eqns. (10)-(11) it is not hard to prove
the following statement [11].

{\bf Proposition}. The representation $T_{\chi}$ is irreducible
if and only if $c_1$ and $c_2$ are not integers  or $c_1$ and $c_2$
coincide with some of the numbers  $l_1,l_2,..,l_{n-1}$.

When both $c_1$ and $c_2$ are integers not coinciding simultaneously
with any two of the integers $l_1,l_2,..,l_{n-1}$, representations
from the 'principal nonunitary series' are no longer irreducible,
and the corresponding irreps are extracted from these reducible
representations (we call them {\it irreps of integer type}).

Two irreps $T_{\chi}$ and $T_{{\chi}'}$ from the Proposition with
$\chi$ and ${\chi}'$ differing only by interchange
$(c_1, c_2)\leftarrow \! \to (c_2, c_1)$, are equivalent.
Periodicity of the function $f(w)=[w]_q$ implies the following:
the representations $T_{\chi}$ and $T_{{\chi}'}$ with
${\chi}'=(l_1,...,l_{n-1};c_1 +{i\pi k\over h},c_2 -{i\pi k\over h})$,
$ k\in {\bf Z}$, are equivalent for $q={\rm exp\ }h,\  h\in {\bf R}$;
the representations $T_{\chi}$ and $T_{{\chi}'}$ with
${\chi}'=(l_1,...,l_{n-1};c_1 +{\pi k\over h},c_2 -{\pi k\over h}),
\ k\in {\bf Z}$, are equivalent for $q={\rm exp\ i}h,\ \ h\in {\bf R}$.
For this reason, we impose the restriction
$0\le {\rm Im}\ c_1<{\pi\over h}$ (respectively the restriction
$0\le {\rm Re}\ c_1<{\pi\over h}$)  in the case of
$q={\rm exp\ }h$ (resp. of $q={\rm exp\ i}h $) ($h\in {\bf R}$ in
both cases).

The classification of irreducible representations of the
algebra  $U_q(u_{n,1})$  is completely analogous to that of the
non-deformed algebra $u(n,1)$ (see e.g. [8, 13]).

For our purposes it will be useful to reproduce here the list of
different classes of irreps of $U_q(u_{n,1})$ which are
'unitary' for $q=e^h,\  h\in {\bf R}$ (the sequence of
numbers $a_1, a_2,..., a_k$ will be called {\it contracted} if
$a_{i-1}-a_i=1$ for $i=2,3,...,k$).
\smallskip

I. Principal continuous series of irreps $T_{\chi }$:  $c_1$
   and $c_2$ are such that $c_1=\bar {c_2}$.
\smallskip
II. Supplementary continuous series of irreps $T_{\chi }$:
    $c_1, c_2 \in {\bf R}$ and, moreover, there exist
    such $l_k$ and $l_s$\ $(k,s=1,2,...,n-1)$ that
    $\vert c_1-l_k\vert <1,\  \vert l_s-c_2\vert <1,$ and
    the sequence $l_k,l_{k+1},...,l_s$, if $c_1 > c_2,$ or
   the sequence $l_s,l_{s+1},...,l_k$,
    if $c_1 < c_2$, is contracted.
\smallskip
III. Strange series of irreps $T_{\chi}$:
     ${\rm Im}\ c_1={\rm Im}\ c_2={\pi \over h}$.

\noindent {For these three continuous 'unitary' series, the irreps}
$T_{\chi}$ when restricted to subalgebra $U_q(u_n)$ contain those
irreps $T_{{\bf l}_n}$ for which the condition (8) is satisfied.
\medskip
Classes of 'unitary' irreps of {\it integer type }\ ($c_1,\ c_2$
\ not both coincide with some of  $l_1,...,l_{n-1}$; we use the
denotion $l_0=\infty , l_n=-\infty )$.
\smallskip
IV. Irreps  $D^{ij}_+(l_1,...,l_{n-1};c_1,c_2)$ and
            $D^{ij}_-(l_1,...,l_{n-1};c_1,c_2)$ where
   $l_{i-1}>c_1>l_i$, $l_{j-1}>c_2>l_j$,\ \ $1\le i\le j\le n$.
   Moreover, either \ $i\!=\!j$ \ holds, or the sequence
   $c_1, l_i, l_{i+1}, ...,l_{j-1}$ for $D^{ij}_+$
  (the sequence $l_i, l_{i+1}, ...,l_{j-1},c_2$ for $D^{ij}_-$)
is contracted. The irrep $D^{ij}_+(l_1,...,l_{n-1};c_1,c_2)$
(resp. irrep $D^{ij}_-(l_1,...,l_{n-1};c_1,c_2)$ ) contains with
unit multiplicity those and only those irreps of $U_q(u_n)$
for which the condition (8) and the conditions $l_{in}>c_1$,
\  $l_{jn}>c_2$ (resp. $l_{in}\le c_1$,\ \ $l_{jn}\le c_2$) are
satisfied.
\smallskip
V. Irreps  $\tilde D^{ij}_+(l_1,...,l_{n-1};c_1,c_2)$ and
            $\tilde D^{ij}_-(l_1,...,l_{n-1};c_1,c_2)$ where
   $c_1=l_i,\ 1\le i\le n-1,$ and $c_2$ is an integer such
   that  $l_{j-1}>c_2>l_j$,\  $1\le j\le n$.
   For  $\tilde D^{ij}_+$, moreover, either $i<j$ and the sequence
   $l_i,l_{i+1},...,l_{j-1},c_2$ is contracted, or $i\ge j$ and
the sequence  $l_j,l_{j+1},...,l_i$ is contracted.
  For $\tilde D^{ij}_-$, either $i<j$ and the sequence
 $l_i,l_{i+1},...,l_{j-1}$ is contracted, or $i\ge j$
and the sequence  $c_2,l_j,l_{j+1},...,l_i$ is contracted.
The  $\tilde D^{ij}_+(l_1,...,l_{n-1};c_1,c_2)$
  (resp. $\tilde D^{ij}_-(l_1,...,l_{n-1};c_1,c_2)$) contains
with unit multiplicity those and only those irreps of $U_q(u_n)$
for which the condition (8) and the condition $l_{jn}>c_2$\
 (resp. $l_{jn}\le c_2$) are satisfied.
\smallskip
VI. Irreps  $D^{i}_+(l_1,...,l_{n-1};c_1,c_2)$ and
            $D^{i}_-(l_1,...,l_{n-1};c_1,c_2)$ where $c_1=c_2=c$
is an integer such that $l_{i-1}>c>l_i$,\ $1\le i\le n$.
The $D^i_+$ (resp. $D^i_-$) contains with unit multiplicity
those and only those irreps of $U_q(u_n)$ for which the
condition (8) and the condition $l_{in}>c$ (resp. $l_{in}\le c$)
is satisfied.
\smallskip
There exist additional equivalence relations between irreps from
different classes IV--VI completely analogous to the equivalence
relations of the non-deformed case (we do not give them here,
see e.g. [8]). Remark that two reducible representations
$T_{\chi}$ and $T_{{\chi}'}$ with $\chi =(l_1,...,l_{n-1};c_1,c_2)$
and $\chi =(l_1,...,l_{n-1};c_2,c_1)$ contain equivalent irreps
(from classes IV-VI) of $U_q(u_{n,1})$.
\medskip
\noindent { At  $q=e^{i h},\  h\in {\bf R}$, the classes I and }
III (with minor modification:
${\rm Re}\ c_1={\rm Re}\ c_2={\pi \over h}$
instead of ${\rm Im}\ c_i$) are the only classes that survive
in the classification of 'unitary' irreps.

It is worth to mention the following: the only class from the
above presented list of irreps which is absent in the classical
limit (disappears at $q\to 1$  or, equivalently, at $h\to 0$) is
the class III (strange series) of 'unitary' irreps.
\bigskip
\centerline {\bf 4. EVALUATION OF BARYON MASSES}
\bigskip
One has to form state vectors for baryons ${1\over 2}^+$ that
constitute {\sl 20}-plet of $U_q(u_4)$ whose decomposition with
respect to $U_q(u_3)$ is
${\sl 20}={\bf 8}+{\bf 3}+{\bf 3^*}+{\bf 6}$. To this end,
we use the embedding $U_q(u_4)\in U_q(u_{4,1})$ and the
(orthonornalized) Gel'fand-Tsetlin basis elements in the form (9)
constructed in accordance with the aforementioned canonical
chain, fixing $n=4$. For instance, for the isodoublet of nucleons
(contained in octet) we have:
$$ \vert N\rangle\leftarrow\!\rightarrow\vert \chi; {\bf l}_4,
{\bf l}_3, {\bf l}_2, l_Q\rangle
$$
where $\chi\equiv (l_1,l_2,l_3;c_1,c_2)$ labels some appropriate
(that is, such that contains the  ${\it 20}$-plet of $U_q(u_4)$)
irrep of the 'dynamical' $U_q(u_{4,1})$;
${\bf l}_4\equiv (p+1,p-1,p-3,p-4)$ and
${\bf l}_3\equiv (p+1,p-1,p-3)$ label ${\it 20}$-plet of
$U_q(u_{4})$ and ${\bf 8}$-plet of its subalgebra $U_q(u_3)$
respectively; ${\bf l}_2$ means isodoublet $(p+1,p-1)$ of
nucleons, and $l_Q$ labels charge states within it. The
quantity ${\bf l}_2$ equal to $(p,p-2)$ characterizes
another ($\Xi$-) isodoublet belonging to octet. State
vectors for the rest of members of the ${\it 20}$-plet are
constructed analogously.
\medskip
Mass operator, according to the concept of pseudounitary
dynamical group [7-8] extended to present ($q$-deformed) case, is
constructed in terms of those "noncompact" generators of
$U_q(u_{4,1})$ which respect the isospin-hypercharge symmetry
$U_q(u_2)$, but break all higher (flavour) symmetries. We take it
in the form
$$\eqalignno {\hat {M_4} = M_o^{(4)} &+ \gamma A_{45}A_{54}
+ \delta  A_{54}A_{45}                             &{}\cr
 &+ \alpha A_{35}\tilde A_{53} +
  \beta \tilde A_{53}A_{35} +\tilde {\alpha} \tilde A_{35} A_{53} +
  \tilde {\beta} A_{53}\tilde A_{35}\ \        &{(12)} }
$$
and put $\alpha =\tilde {\alpha },\  \beta =\tilde {\beta }$
in order to reduce the number of independent parameters.
As a result of calculation of the matrix elements
$\langle N\vert\ \hat {M_4}\ \vert N\rangle\equiv m_N$,\
$\langle \Xi\vert\ \hat {M_4}\ \vert\Xi\rangle\equiv m_{\Xi}$,
etc., within the representation
$D^{12}_+(p-1,p-3,p-4;p,p-2)$\ ($p$ being an arbitrary
fixed integer), we obtain the following expressions for masses of
baryons ${1\over 2}^+$ belonging to ${\sl 20}$-plet of $U_q(u_4)$.
\smallskip
\noindent {(i) Octet baryons:}
$$\eqalignno {
m_N &= M_{\bf 8}+{[2][3]\over [6]}([5]\alpha +\beta),          &{}\cr
m_{\Lambda } &= M_{\bf 8} +
{1\over [6]}\Bigl ({[5][4]^2\over [2]^2} + [2]^2\Bigr )\alpha +
{[2]^2[3]\over [6]}\beta ,                                     &{}\cr
m_{\Xi } &= M_{\bf 8} + {1\over [6]}\Bigl ([2]([5]+[3])
+{[5][3]\over [2]}([5]-[3]-2)\Bigr )\alpha +
{[2]^2[4]\over [6]}\beta,                                     &{}\cr
m_{\Sigma }&= M_{\bf 8}+{[4]^2[5]\over [2]^2[6]}\alpha
+{[2][4]\over [6]}\beta ;                                     &(13)}
$$
(ii) triplet baryons:
$$\eqalignno {
m_{{\Xi }_{cc}} &= M_{\bf 3}+{[2][3][5]\over [6]}\alpha
+{[2]\over [6]}\Bigl (([4]-[2])^2-1\Bigr )\beta ,   &{}\cr
m_{{\Omega }_{cc}} &= M_{\bf 3}+{[2]^2[5]\over [6]}\alpha
+{[2][4]\over [6]}([3]-2)\beta ;                    &(14) }
$$
(iii) antitriplet baryons:
$$\eqalignno {
m_{{\Xi }_c'} &= M_{\bf 3^*}+{[2][3][5]\over [6]}\alpha
+{[2][3]\over [6]}([3]-2)\beta ,                   &{}\cr
m_{{\Lambda }_c} &= M_{\bf 3^*}+{[3]\over [6]}([3]-2)([5]+2)\alpha
+{[3]^2\over [6]}([3]-2)\beta ;                    &(15)  }
$$
(iv) sextet baryons:
$$\eqalignno {
m_{{\Sigma }_c} &= M_{\bf 6}+{[2][3][5]\over [6]}\alpha
+{[2]\over [6]}\Bigl (1+([3]-1)([3]-2)\Bigr )\beta,   &{}\cr
m_{{\Xi }_c} &= M_{\bf 6}+{[3]+2([3]-1)[5]\over [6]}\alpha
+{2[5] -[3]+4\over [6]}\beta,                  &{}\cr
m_{{\Omega }^0_c} &= M_{\bf 6}+
                 {[2]\over [6]}\Bigl ([3]+([3]-2)[5]\Bigr )\alpha
+{[2]^2[4]\over [6]}\beta .               &(16) }
$$
In the above expressions for masses, we have used the notations
$$\eqalign { &M_{\bf 8}=M_0+{[3]\over [6]}([5]\gamma +\delta ) , \cr
	     &M_{\bf 3}=M_0+{[2][4]\over [6]}(\gamma +\delta ) , \cr
	     &M_{\bf 3^*}=M_0+{[3]\over [6]}([4]\gamma +[2]\delta ) , \cr
	     &M_{\bf 6}=M_0+{1\over [6]}([2][5]\gamma +[4]\delta ) . }
$$
As it is seen, the integer $p$ does not enter the obtained
expressions for baryon masses. In other words, the approach
which we follow here is insensitive to the substitution
$U_q(su_{4,1})\rightarrow U_q(u_{4,1})$.

Let us check now that in the 'classical' limit $q\to 1$ octet
masses satisfy GMO-relation.
Indeed, for $q=1$ we have that $m_N=\tilde M_{\bf 8}
+5\alpha +\beta$,\ \ $m_{\Lambda }=\tilde M_{\bf 8}+4\alpha +2\beta $,
\ \ $m_{\Xi }=\tilde M_{\bf 8}+{8\over 3}(\alpha +\beta )$,
\ \ $m_{\Sigma }=\tilde M_{\bf 8}+{10\over 3}\alpha +{4\over 3}\beta $
(here $\tilde M_{\bf 8}=M_0+{15\over 6}\gamma +{1\over 2}\delta$),
and the relation (1) is obviously satisfied.
\bigskip
\centerline {\bf 5. $q$-ANALOG OF BARYON OCTET MASS RELATION}
\bigskip
We are predominantly interested in an octet MR with $q$-dependent
coefficients. From eqns. (13), through the differences
$[2]m_N -m_{\Lambda},\ [2]m_{\Sigma}-m_{\Xi},\ $ and
\ $[2](m_{\Sigma}-m_N)+m_{\Lambda}-m_{\Xi},\ $ it is straightforward
to find the formulas which express independent parameters $M_{\bf 8},\
 \alpha$\  and  $\beta$\  in terms of baryon masses:
$$\eqalign {\alpha &=
{[2](m_{\Sigma}-m_N)+m_{\Lambda}-m_{\Xi}
               \over [3]\{[3]+[2]^2([4]-[2])-3[5]\} } ,          \cr
    \beta &= {[6]\over [2]^2}(m_{\Lambda}-m_{\Sigma}) - \alpha , \cr
M_{\bf 8} &= {[2]m_{\Sigma}-m_{\Xi}\over [2]-1}
            -{[2]^2\over [2]-1} \alpha .         }$$
Substituting these expressions into the last relation of eq.(13),
we arrive at the desired $q$-analog MR, namely,
$$\eqalignno { [2]m_N+{[2]\over [2]-1}m_{\Xi }&=[3]m_{\Lambda }
+\Bigl ({[2]^2\over [2]-1}-[3]\Bigr )m_{\Sigma }         &{}\cr
        &+{A_q\over B_q}\bigl ([2]m_N+m_{\Xi }
                    -m_{\Lambda }-[2]m_{\Sigma }\bigr ) &{(17)}  }
$$
where
$$\eqalignno {A_q &=[2]^4+[2]^3\bigl ([5]-[4]\bigr )
                   +[2]^2\bigl ([6]-[5]\bigr )-
                    [2]\bigl ([6]+[4]^2\bigr )+[4]^2,     &{(18)}\cr
              B_q &=\Bigl ([2]^3-[2]^2[4]+3[5]-
                    [3]\Bigr )([2]-1).                   &{(19)}  }
$$
The relation (17) constitutes our main result. Here we observe
nontriviality of the coefficients at $m_{\Xi }$ and $m_{\Sigma }$
and, as most unexpected thing, the appearance of that additional
structure (second line in (17)) with $A_q/B_q$ as its coefficient.

Strictly speaking, the relation just obtained is not
a mass relation. However, at any fixed value of $q$ it yields some
'candidate' MSR. For this reason the $q$-analog relation (17) may be
viewed as a continuum of candidate MSR's for baryon masses, only
few of which may be considered as realistic mass sum rules.
\bigskip
\centerline {\bf 6. DEFINING $q$-POLYNOMIAL AND A NEW BARYON MASS SUM RULE}
\bigskip
Now the problem consists
in finding the value(s) of deformation parameter
at which the relation (17) yields most realistic MSR(s). Clearly, a
straightforward way to proceed would be to insert the empirical data
for the octet baryon masses into (17) and then solve the equation
with respect to $q$. However, we think of this way as not the best one
for two reasons: (i) the equation for $q$ this way appears to be rather
complicated; (ii) so obtained values of deformation parameter
would be neccessarily {\it 'non-rigid'} ones reflecting
approximate procedure of solving the equation as well as
errors of experimental data and averaging over isomultiplets.
Fortunately, there exists another approach which is somewhat
analogous to reasonings used in [5,6] for the case of vector
mesons and which leads to {\it 'rigidly fixed'} values of $q$.
To this end, let us return again to the 'classical' case.
As already mentioned, the value $q=1$ must result in the
standard GMO-relation (a kind of the
'correspondence principle').
Indeed, $A_{q=1}=0\ $, ($B_{q=1}\ne 0$), and
$m_N+m_{\Xi}={3\over 2}m_{\Lambda}+{1\over 2}m_{\Sigma }$ results.
But now we observe the point of oversimplification (due to vanishing
of $A_q$ at $q=1$) when reducing the $q$-analog to GMO-relation.
Adopting this as {\it a hint of how to search other candidate
values of} $q$, we proceed by rewrighting the $A_q$ (which is
7-th order polynomial in $q$-deuce $[2]_q$) in its
completely factorized form:
$$\eqalignno   { A_q&=([2]-2)[2]^3([4]-[2])           &{}\cr
		    &=([2]-2)[2]^4([2]^2-3)           &{(18')}  }
$$
(the recursion $[n]_q=[2]_q[n-1]_q-[n-2]_q$ for $q$-numbers
is useful in doing this). Since the 'classical' GMO-relation
corresponds to vanishing of $A_q$ because of the fact that
\medskip
(i) $([2]-2)=0$,
\medskip
\noindent {it is natural to examine } the remaining cases
when $A_q$ turns into zero:
\medskip
(ii) $[2]=0$;
\medskip
(iii) $[4]-[2]\equiv [2]([3]-2)\equiv [2]([2]^2-3)=0,\ \ [2]\ne 0.$
\medskip
\noindent {The case (ii) leads to the MR}\ \ $m_{\Lambda } =
m_{\Sigma }$\ \ which is not very good since, with empirical data,
its accuracy is $\approx 6.5\%$). It is interesting, nevertheless,
to compare this case with the second nonet Okubo's formula
$m_{\rho }=m_{\omega }$ (which was shown to follow from the $q$-analog
of vector meson MR,\ see [5],\ just if the same restriction $[2]_q=0$
has been applied). Remark that in both meson and baryon cases,
these MSR's relate isosinglet and isotriplet masses.
\medskip
Now let us consider the most interesting case (iii), that is,
the values of $q$ that solve $[2]_q=\pm \sqrt{3}$. These are
respectively $q_+=e^{i\pi \over 6}$ and $q_-=e^{i5\pi \over 6}$.
At these values, $[4]_{q_\pm} = \pm \sqrt{3}$ (that is, both $q_+$
and $q_-$ solve the equation $[4]_q-[2]_q=0$) and also $[3]_{q_\pm} = 2$
($q_+$ and $q_-$ both solve the equation $[3]_q-2=0$).

Two candidate MSR's follow from (17) at $q_+,\ q_-$:\ the relation
$$m_N + {1\!+\!\sqrt 3\over 2}\ m_{\Xi} = {2\over{\sqrt 3}}\ m_{\Lambda }
	 + {9\!-\!\sqrt 3\over 6}\ m_{\Sigma }        \eqno (20)
$$
and the relation
$$m_N + {1\!-\!\sqrt 3\over 2}\ m_{\Xi} = {-2\over{\sqrt 3}}\ m_{\Lambda }
	 + {9\!+\!\sqrt 3\over 6}\ m_{\Sigma }        \eqno (20')
$$
respectively. The second candidate MSR (which corresponds to $q_-$)
shows bad agreement with data. Fixing $q=q_+$,\ however, we get
a {\it surprizingly good mass relation}: with empirical
values [14] for octet ${1\over 2}^+$ baryon masses
($m_N=938.9\ MeV$,\ \  $m_{\Xi}=1318.1\ MeV$,\ \ $m_{\Lambda }=
1115.6\ MeV$,\ and $m_{\Sigma }=1193.1\ MeV$) we have
$2739.5\ MeV \approx 2733.4\ MeV$. That is,
eq.(20) holds with $0.22 \%$ accuracy! For comparison recall
that usual GMO-relation (1) is satisfied within $0.57 \%$.

Strictly speaking, the case of $q\rightarrow q_+$\  essentially
differs from the classical ($q=1$) situation, besides
irrationality of the coefficients in (20), yet by the following
peculiarity. All the nasses in (13)-(16) become infinite
in this case because of pole singularity, since
$[6]\equiv [2][3]([3]-2)\rightarrow 0$\ when  $q$ tends to $q_+$.
\ In order to circumvent this difficulty, one has first to interprete
the invariant (background for the ${\sl 20}$-plet) mass $M_0$\ and all
the masses $m_{B_i}$ (with $B_i$ running over the set of
baryon symbols in formulas (13)-(16) ) as infinite 'bare' masses,
then going over to finite 'physical' masses by making use of a
multiplicative renormalization through the substitution
$M_0\rightarrow {M_{0 ({\rm phys.})}\over [6]}$,
\ $m_{B_i}\rightarrow {m_{B_i ({\rm phys.})}\over [6]}$. Let us
remark that such a substitution does not affect the explicit form
of the $q$-analog mass relation (17) and the MSR (20).
\bigskip
\centerline {\bf 7. QUANTUM ALGEBRA $U_q(u_{4,1})$\ REMOVES 'DEGENERACY'}
\bigskip
Calculations analogous to those of section 5 were performed also
for another specific representation of dynamical algebra. Namely,
with the {\sl 20}-plet  $(p+1,p-1,p-3,p-4)$\  of $U_q(u_4)$
embedded into the (integer type) irrep
${\tilde D}^{32}_+(p-1,p-3,p-4;p-4,p-2)$\ of $U_q(u_{4,1})$,
we have obtained the expressions for baryon masses,
$$\eqalignno {
m_N &= {M_{\bf 8}}'+{[2][3]\over [6]}(\alpha +[5]\beta),    &{}\cr
   m_{\Lambda } &={M_{\bf 8}}' +
{[3]\over [6]}([3]^2+[3]-4)\alpha
+{[2]^2[3][5]\over [6]}\beta ,               &{}\cr
   m_{\Xi } &={M_{\bf 8}}' +
{[2]\over [6]}([3]^3-4[3]+1)\alpha +
{[2]^2[4][5]\over [6]}\beta,                  &{}\cr
   m_{\Sigma }&={M_{\bf 8}}' +
{([3]-1)^2\over [6]}\alpha +{[2][4][5]\over [6]}\beta ,     &{(21)}  }
$$
(here ${M_{\bf 8}}'\equiv M_0+{[3]\over [6]}\gamma +{[3][5]\over [6]}\delta $),
as well as the following $q$-analog of octet MR
$$  m_N-m_{\Lambda }-{[2]-1\over [2]^2}M_C+
{[3]([2]-1)\over [2]^2}M_D=
{\tilde A_q\over \tilde B_q}M_C         \eqno (22)
$$
where
$$\eqalign { M_C&=[2](m_N-m_{\Sigma })+m_{\Xi }-m_{\Lambda }, \cr
M_D&=([2]+1)m_{\Lambda }-[2]m_N-m_{\Xi },  \cr
\tilde B_q&=[2][3]^2-[3]^2-[2][3]-2[2]+5, }
$$
and the defining polynomial is of the form
$$\tilde A_q=([2]-2)([3]^2-5).   \eqno (23)
$$
Besides the 'classical' root $[2]_q=2$ (equivalent to $q=1$)
which determines the standard GMO-relation (1),
the polynomial $\tilde A_q$ has four roots more, corresponding
to $[3]_q=\pm \sqrt 5$. For instance, with $[2]_q=\sqrt {1+\sqrt 5}$
(this choice provides $[3]_q=\sqrt 5$) we have the MSR
$$\eqalignno { 2\Bigl (\sqrt 5-\sqrt {\sqrt 5 +1}\Bigr )m_N +
2\Bigl (\sqrt {\sqrt 5 +1}-1\Bigr )&m_{\Xi}=     {}\cr
\Bigl (2\sqrt 5 -4+\sqrt {\sqrt 5 -1}\Bigr )&m_{\Lambda }+
\Bigl (2-\sqrt {\sqrt 5 -1}\Bigr )m_{\Sigma }.     &{(24)}  }
$$
Examination of this MSR with inserted empirical masses shows that
it is not very satisfactory (the accuracy is $\approx 2.7\%$).
Nevertheless, one could conclude from this second example
that distinct dynamical representations may yield essentially
different $q$-analogs of octet MSR, with different
defining $q$-polynomials. While the presence of factor $[2]_q-2$
is common feature of all defining polynomials, the difference would
lie in sets of extra roots (compare $[2]_q=0,\ [2]_q=\pm\sqrt 3$ of
the polynomial ($31'$) and the roots $[2]_q^2=1\pm\sqrt 5$  of
$\tilde A_q$ in this section).
Since when applying classical Lie algebra $u(4,1)$
{\it all the representations } yield [7] the GMO-relation
and nothing else (a kind of 'degeneracy'), we could say that
application of dynamical $q$-algebra $U_q(u_{4,1})$,
for octet baryon mass relations, removes this degeneracy.

It turns out, however, that by processing the expressions (21)
in some another way we arrive at the $q$-analog of octet MR which
coincides with the first $q$-MR, eq.(17), in all the
points (coefficients at masses, the combination of masses in curly
brackets and, most inportant, the defining polynomial $A_q$
in front of the $M_C$ in (17) ) except for the change
$B_q\rightarrow \tilde B_q$ of the $q$-polynomial in denominator
(it plays no essential role in reducing to MSR's (1) and (20) ).
Anyway, we conclude that application of the $q$-algebra
$U_q(u_{4,1})$\ as dynamical one 'removes degeneracy' at least
in the sense that in the framework of the quantum algebra we get,
besides eq.(1), mass sum rules of novel type (including such
accurate one as eq.(20) ). Those novel MSR's seem to reflect
some interesting hadronic dynamics, and this certainly
deserves further study.
\bigskip
\centerline {\bf 8. CONCLUSIONS AND OUTLOOK}
\bigskip
Extending the approach of dynamical (pseudo)unitary groups
to the quantum groups $U_q(su_n)$\ ($n$-flavor symmetry) and
$U_q(u_{n,1})$\ ('dynamical' symmetry) we have obtained at $n=4$
the $q$-dependent expressions for masses of baryons ${1\over 2}^+$
from which $q$-MR's follow.

The $q$-deformed relation (17) is of interest, first of all,
as a 'continuum' of possible MSR's and also due to the fact that
it contains unusual mass-dependent term
with $A_q/B_q$ in front of it. It is just this point where
the concept of {\it defining $q$-polynomial} arises:
the polynomial $A_q$, by means of its roots, determines concrete
octet baryon MSR's (namely, the classical GMO-relation (1) and
this novel, very accurate, relation (20) with $\sqrt 3$ contained
in some of its coefficients, as most successful ones; the MSR ($20'$)
and the relation  $m_{\Lambda } = m_{\Sigma }$,\ as less successful
MSR's).

It is important to stress once more that, {\it due to its irrational
coefficients}, the relation (20) is of unconventional nature. This
property is obviously connected with the root of unity
$q_+=e^{i\pi \over 6}$\ and probably reflects some 'nonperturbative'
(topological) information encapsulated in the model under
consideration at such value of deformation parameter.

To make last assertion somewhat more transparent, it is useful
to present the $q$-MR (17) in the form
$$       (1-\Delta_N)m_N + (1+\Delta_{\Xi })m_{\Xi} =
 \Bigl ({3\over 2}-\Delta_{\Lambda }\Bigr )m_{\Lambda }
     + \Bigl ({1\over 2}+\Delta_{\Sigma }\Bigr )m_{\Sigma },   \eqno (25)
$$
\noindent {where}
$$\eqalignno {
  &\Delta_N\equiv {A_q\over B_q}, \hskip 95 pt
 \Delta_{\Xi }\equiv - {A_q\over [2]B_q} + {1\over [2]-1} - 1,   &{}\cr
&\Delta_{\Lambda }\equiv {A_q\over [2]B_q}+
           {3\over 2}-{[3]\over [2]},   \hskip 40 pt
\Delta_{\Sigma }\equiv - {A_q\over B_q}+{[2]\over [2]-1}
                       - {[3]\over [2]} - {1\over 2}.           &{(26)} }
$$
\noindent {Since} $A_q=0$ at $q=1$, we have that all $\Delta_k$
(here $k$ takes the 'values' $N,\ \Xi ,\ \Lambda $ and $\Sigma $)
equal zero in the classical limit. Therefore, it is natural
to consider the quantities $\Delta_k$  at $q\ne 1$ as {\it 'corrections'
to the classical coefficients} due to $q$-deformation. Obviously,
at values of $q$ which are very close to unity these corrections
do not deviate substantially from zero.
\smallskip
At $q=q_+$ again $A_q$ vanishes. However,  now  all $\Delta_k$
other than  $\Delta_N$  are not small ('perturbative') quantities,
but become of the order of magnitude
comparable with the corresponding classical coefficients
(for instance, $\Delta_{\Sigma }=
   {\sqrt 3\over \sqrt 3 -1}-{2\over \sqrt 3}-{1\over 2}\approx 0.71$
to be compared with the coefficient ${1\over 2}$ in
classical MSR (1)).

Of course, further study is needed in order to clarify, among others,
such issues as: (in)dependence of the results, concerning
$q$-analog MR's for baryons ${1\over 2}^+$ and the appearance of
new resulting mass sum rules (like the MSR (20)), on the choice of
dynamical representation of the quantum algebra $U_q(u_{4,1})$;
the issue of reducibility of the infinite dimensional
representations used within our approach at $q$ being specific
roots of unity, as well as details of dynamics at those values
of the deformation parameter; the question about possibility
to associate with baryons some topological structures (knots or
links) in a manner more or less similar to the treatment
in the case of vector mesons [6]. We hope to analyze
these problems in subsequent publications.

\medskip
This research was partially supported by the International
Science Foundation under the Grant U4J000 and by the Ukrainian
State Foundation for Fundamental Research.
\bigskip
\centerline {\bf REFERENCES}
\bigskip
\item {1.}Iwao S 1990 {\sl Progr. Theor. Phys.} {\bf 83} 363.
\item {2.}Raychev P P, Roussev R P and Smirnov Yu F 1990
      {\sl J. Phys. G} {\bf 16} 137.
\item {}Bonatsos D et al. 1990 {\sl Phys. Lett.} {\bf 251B} 477.
\item {3.}Celeghini E et al. 1991 {\sl Firenze preprint} DFF 151/11/91.
\item {4.}Biedenharn L C 1990 {\sl An overview of quantum groups}, in:
      Proc. $XVII^{th}$ Int. Coll. on Group Theor. Methods in Physics
      (Moscow, June 1990).
\item {}Zachos C 1991 {\sl Paradigms of quantum algebras}, in:
      Symmetry in Science V (B Gruber, L C Biedenharn, and
      H D Doebner, eds.), pp 593-609.
\item {}Kibler M 1993 {\sl Lyon preprint} LYCEN/9358.
\item {5.}Gavrilik A M 1993 , in: Physics in Ukraine. {\sl Elementary
      Particle Physics and Quantum Field Theory}
      (Proc. Internat. Conference, Kiev), pp 38-41;
      Gavrilik A M, Tertychnyj A V {\sl Kiev preprint} ITP-93-19E.
\item {6.}Gavrilik A M 1994 {\sl J. Phys. A} {\bf 27} L91;
           {\sl Kiev preprint} ITP-93-28E.
\item {7.}Gavrilik A M, Shirokov V A 1978 {\sl Yadernaya Fizika}
      {\bf 28} 199.
\item {}Kalman C S 1978 {\sl Lett. Nuovo Cimento} {\bf 21} 291.
\item {8.}Gavrilik A M, Klimyk A U 1989 {\sl Symposia Mathematica}
      {\bf 31} 127.
\item {9.}Novozhilov Yu V 1975 {\sl Introduction to elementary
       particle theory}  (New York, Pergamon).
\item {}Gasiorowicz S 1966 {\sl Elementary particle theory}
      (New York, J Willey).
\item {10.}Jimbo M 1985 {\sl Lett. Math. Phys.} {\bf 10} 63.
\item {11.}Klimyk A U, Groza V A  {\sl Kiev preprint} ITP-89-37R.
\item {}Klimyk A U {\sl Infinite dimensional representations
      of quantum algebras}, to be published.
\item {12.}Chakrabarti A 1991 {\sl J. Math. Phys.} {\bf 32} 1227.
\item {13.}Klimyk A U, Gavrilik A M 1979 {\sl J. Math. Phys.}
      {\bf 20} 1624.
\item {14.}Particle Data Group 1990 {\sl Phys. Lett. B} {\bf 239} 1.

\end